\documentclass[prd,nofootinbib,preprint,superscriptaddress,letter]{revtex4}
\pdfoutput=1

\usepackage{float}
\usepackage{physics,slashed}
\usepackage{amsfonts,amsmath,amssymb}
\usepackage{mathrsfs}
\usepackage{bm}

\usepackage{epsfig}
\usepackage{graphicx}   
\usepackage{subfigure}  
\usepackage{verbatim}   
\usepackage{color}      
\usepackage{hyperref}   
\usepackage{url}
\definecolor{nicecol1}{rgb}{0.56,0.,1.}
\definecolor{nicecol2}{rgb}{1.,0.3,0.8}
\definecolor{nicecol3}{rgb}{0.,1.,0.}
\hypersetup{colorlinks,citecolor=nicecol1,linkcolor=nicecol2,urlcolor=nicecol3}
\usepackage{array}
\usepackage{dcolumn}
\usepackage{multirow}
\usepackage{cprotect}
\usepackage{framed}
\usepackage{setspace}
\usepackage{enumitem}
\setlist{nolistsep}

\usepackage{tikz-feynman,contour}
\tikzfeynmanset{compat=1.1.0}
\usepackage{tikzsymbols}

\renewcommand\[{\begin{equation}}
\renewcommand\]{\end{equation}}

\begin{document}

\title{Dynamical Inflation Stimulated Cogenesis}
\author{Debasish Borah}
\email{dborah@iitg.ac.in}
\affiliation{Department of Physics, Indian Institute of Technology Guwahati, Assam 781039, India}
\author{Arnab Dasgupta}
\email{arnabdasgupta@pitt.edu}
\affiliation{Pittsburgh Particle Physics, Astrophysics, and Cosmology Center, Department of Physics and Astronomy, University of Pittsburgh, Pittsburgh, PA 15206, USA}
\author{Daniel Stolarski}
\email{stolar@physics.carleton.ca}
\affiliation{Department of Physics, Carleton University
1125 Colonel By Drive, Ottawa, Ontario K1S 5B6, Canada}
\affiliation{Physics Department 0319, University of California San Diego,
9500 Gilman Drive, La Jolla, CA 92093-0319, USA\vspace{1cm}}

\begin{abstract}
\vspace{.5cm}
\hspace{-0.6cm} We propose a minimal setup that realises dynamical inflection point inflation, and, using the same field content, generates neutrino masses, a baryon asymmetry of the universe, and dark matter.  A dark $SU(2)_D$ gauge sector with a dark scalar doublet playing the role of inflaton is considered along with several doublet and singlet fermions sufficient to realise multiple inflection points in the inflaton potential. The singlet fermions couple to SM leptons and generate neutrino masses via the inverse seesaw mechanism. Those fermions also decay asymmetrically and out of equilibrium, generating a baryon asymmetry via leptogenesis. Some of the fermion doublets are dark matter, and they are produced via inflaton decay and freeze-in annihilation of the same fermions that generate the lepton asymmetry. Reheating, leptogenesis, and dark matter are all at the TeV scale. 
\end{abstract}

\maketitle

\section{Introduction}

The cosmic microwave background (CMB) observations show that our universe is homogeneous and isotropic
on large scales up to an impressive accuracy \cite{Planck:2018vyg, Planck:2018jri}. Such observations lead to the so called horizon and flatness problems which remain unexplained in the description of standard cosmology. The theory of cosmic inflation that posits a phase of  rapid accelerated expansion in the very early universe was proposed in order to alleviate these problems \cite{Guth:1980zm,Starobinsky:1980te,Linde:1981mu}. While there are several viable inflationary models discussed in the literature, not all of them are motivated from particle physics point of view. Here we consider the approach where the inflation potential arises from Coleman-Weinberg corrections to scalar potential in a well-motivated particle physics setup \cite{Linde:1981mu, Kallosh:2019jnl, Guth:1982ec, Shafi:2006cs, Badziak:2008gv, Enqvist:2010vd, Okada:2014lxa, Okada:2016ssd, Choi:2016eif, Okada:2017cvy, Urbano:2019ohp, Bai:2020zil, Ghoshal:2022hyc, Ghoshal:2022jeo, Ghoshal:2022zwu} naturally leading to a low scale inflation. In particular, we build on dynamical inflection point inflation~\cite{Bai:2020zil} which has an inflection point in the inflaton potential due to the vanishing of the scalar quartic $\beta$-function. This zero easily occurs if the inflaton has both gauge and Yukawa couplings. 

The Standard Model (SM) of particle physics, while extremely successful, has several shortcomings that make it unable to describe our universe. In particular, the minimal renormalizable Standard Model cannot accommodate neutrino masses, does not have a dark matter (DM) candidate, and cannot give rise to a sufficiently large baryon asymmetry of the universe (BAU). All three of these are now very well established and require physics beyond the Standard Model. In this work we use the ingredients \textit{already provided} by the dynamical inflection point inflation scenario to solve all of these problems in a unified framework.


Our model consists of a dark $SU(2)_D$ gauge sector with several vectorlike fermion doublets and singlets. There is a doublet scalar responsible for spontaneous symmetry breaking of the dark gauge symmetry that also plays the role of inflaton. The number of dark sector fields not only dictates the shape of the inflaton potential via Coleman-Weinberg corrections, but can also give rise to dark matter and baryon asymmetry of the universe. The fermion singlets couple to some doublet fermions via the inflaton field to become massive and decouple from the beta functions at low energy. The singlet fields can also couple to the lepton doublets via the SM Higgs doublet, which after the introduction of a bare Majorana mass term for dark sector fermions, can lead to the origin of light neutrino masses via the inverse seesaw mechanism. Interestingly, this bare Majorana mass term of dark sector fermions has a strict upper bound from the requirement of successful Coleman-Weinberg inflation, thereby connecting the seesaw scale with inflationary dynamics. While the heavy dark sector fermions dictate the origin of neutrino mass and BAU via leptogenesis \cite{Fukugita:1986hr}, the remaining dark sector fermions which do not couple to the inflaton field can remain light. They do  contribute to the $\beta$-functions down to their mass threshold and also make up the dark matter of the universe. Due to the relative heaviness of the dark $SU(2)$ gauge boson that provides the portal from the dark matter to the right handed neutrinos (RHN), the relic abundance of DM is generated via the freeze-in mechanism~\cite{Hall:2009bx}. Finally, the SM Higgs also gets its mass from the inflaton vacuum expectation value (vev). 

This paper is organised as follows. In section~\ref{sec:Model}, we introduce the model and detail the inflationary dynamics. In section~\ref{sec:neutrino}, we briefly discuss the origin of light neutrino masses via the inverse seesaw mechanism. In section~\ref{sec:lepto} we discuss the details of cogenesis of the baryon asymmetry of the universe and the dark matter abundance. We conclude in section~\ref{sec:conclude}, and we review the standard inflation formulae that we have used in appendix~\ref{app:review}.

\section{The Model}
\label{sec:Model}

In this section we review the mechanism of dynamical inflection point inflation and describe the ingredients of the model. 
We seek an inflaton potential that can arise from a field theory, has sub-Planckian field excursion, and satisfies the constraints from observations of the CMB. Our starting point is inspired by the Coleman-Weinberg potential~\cite{Coleman:1973jx}, 
\begin{equation}
V(\Phi) \sim \lambda(\Phi)\Phi^4,
\label{eq:schematic}
\end{equation}
which has long been studied as a possible inflation model~\cite{Linde:1981mu}. This potential has a plateau that easily gives rise to slow roll as well as a global minimum that the field will naturally roll towards. 

For sub-Planckian field excursions, the scalar spectral index $n_s$ is controlled by the second derivative of the potential (see Eqs.~\eqref{eq:slowRollParams} and~\eqref{eq:inflationObservables}). For a single scalar field we have $\lambda(\Phi) \sim \log(\Phi/f)$, but such a model gives values of $n_s$ that are smaller than the observed values~\cite{Kallosh:2019jnl}. Therefore, one must engineer a smaller second derivative at the point in field space where the cosmological scales leave the horizon. This naturally occurs in potentials where the second derivative can vanish, namely those that have an inflection point. Potentials of the type in Eq.~\eqref{eq:schematic} have inflection points if the $\beta$-function for the quartic coupling, $\beta_\lambda$, has a zero~\cite{Bai:2020zil}. This can occur if $\Phi$ has a gauge charge and couples to fermions. In order to get a suitable inflaton potential, one needs two zeros in $\beta_\lambda$ that are parameterically separated. This ensures that the inflection point is sufficiently far from the minimum so that there can be 60 e-folds of inflation.

Given the above considerations, the field content for the model is shown in Table~\ref{tab:model}, which is very similar to the model in~\cite{Bai:2020zil}. There is a dark $SU(2)$ gauge symmetry, and we impose a global $Z_3\times Z_2$ symmetry.\footnote{$Z_3\times Z_2$ is isomorphic to $Z_6$, but because of the breaking pattern, using the description of the former is simpler.} The $Z_2$ symmetry ensures stability of the dark matter candidate, and we will add small soft breaking of the $Z_3$ in section~\ref{sec:neutrino}. Here, $\Phi$ plays the role of the inflaton, the $\psi_i$ are dark matter, and linear combinations of the $\psi_1$ and $\chi$ are the right handed neutrinos that give neutrinos mass and generate the observed baryon asymmetry via leptogenesis.

\begin{table}[!h]
    \centering
    \begin{tabular}{|c|c|c|c|c|}
    \hline   & $(2s+1)$  & $n_f$ & $SU(2)_D$ & $Z_3\times Z_2$  \\
    \hline    $\psi_{iL}$& 2 & $N_\psi$  & \textbf{2} & $(1,-1)$ \\
    $\psi_{iR}$& 2 & $N_\psi$  & \textbf{2} & $(1,-1)$\\
    $\psi_{1L}$& 2 & 1  & \textbf{2} & $(\omega,1)$\\
    $\psi_{1R}$& 2 & 1  & \textbf{2} & $(\omega^2,1)$ \\
    $\chi_{iL}$& 2 & 2 & \textbf{1} & $(\omega^2,1)$ \\
    $\chi_{iR}$& 2 & 2 & \textbf{1} & $(\omega,1)$ \\
    $\Phi$& 1 & 1 & \textbf{2} &  $(1,1)$ \\
    \hline
    \end{tabular}
    \caption{Particle content beyond the SM. Here $s$ is the particle spin, $n_f$ is the number of flavors, and all fermions are written in Weyl notation. 
    The factor $\omega=e^{2i\pi/3}$ is used to represent 
    $Z_3$ group elements.
    }
    \label{tab:model}
\end{table}


The Lagrangian for the dark sector necessary for the inflection point inflation is given as:
\begin{widetext}
\begin{align}
    -\mathcal{L}_\text{Dark} &\supset
   \left(\sum^{2}_{i=1} y_{iA} \,\overline{\psi_{1L}}\, \Phi \,\chi_{iR} + 
   \sum^{2}_{i=1} y_{iB}\, \overline{\psi_{1L}}\, \widetilde{\Phi}\, \chi_{iR} + 
   \sum^{2}_{i=1} y^\prime_{iA} \,\overline{\psi_{1L}}\, \Phi \,\chi^c_{iL} + 
   \sum^{2}_{i=1} y^\prime_{iB}\, \overline{\psi_{1L}}\, \widetilde{\Phi}\, \chi^c_{iL}  \right. \nonumber \\
   &+\sum^{2}_{i=1} \Tilde{y}_{iA} \,\overline{\chi_{iL}}\, \Phi^\dagger \,\psi_{1R} + 
   \sum^{2}_{i=1} \Tilde{y}_{iB}\, \overline{\chi_{iL}}\, \widetilde{\Phi^*}\, \psi_{1R} + 
   \sum^{2}_{i=1} \Tilde{y}^\prime_{iA} \,\overline{\chi_{iL}}\, \Phi^\dagger \,\psi^c_{1L} + 
   \sum^{2}_{i=1} \Tilde{y}^\prime_{iB}\, \overline{\chi_{iL}}\, \widetilde{\Phi^*}\, \psi^c_{1L}\nonumber \\
   &+ 
\left. \sum^{N_{\psi}+1}_{i=2} M \,\overline{\psi_{iL}}\,\psi_{iR} + \text{h.c.}\right)  + \frac{\lambda_\Phi}{4} |\Phi|^4  
    + \frac{\lambda_H}{4} |H|^4 
    + \lambda_{\Phi H} |\Phi|^2|H|^2 ,
    \label{eq:Lag1}
\end{align}
\end{widetext}
%
where $\widetilde{\Phi} = i \sigma^2 \Phi^*$ and $\widetilde{\Phi^*} = i \Phi^\dagger \sigma^2$. This is the most general renormalizable Lagrangian for the dark sector fields consistent with the global and gauge symmetries.
The Coleman-Weinberg-type inflation scenario requires that the mass term for $\Phi$ be absent or very small, and we have thus not written it in Eq.~\eqref{eq:Lag1}.
In the same spirit, we also do not write a bare mass term for the Higgs. 
The only hard mass term is that for the $\psi_i$, the dark matter states. For simplicity we assume that $M$ is proportional to the identity in $\psi_i$ flavour space. 

Mass terms for $\psi_1$ and $\chi_i$ are forbidden by the $Z_3$ symmetry. Those fields do have Yukawa couplings to the inflaton $\Phi$. The $\psi_{1L}$ and $\psi_{1R}$ are two Weyl doublets, while there are two flavours of $\chi_{iL}$ and $\chi_{iR}$.
Therefore, when $\Phi$ settles to its non-zero minimum, these states will form four Dirac fermions whose mass is proportional to $\langle \Phi \rangle$. To simplify the analysis, we take the following ansatz for the 16 different Yukawa couplings:
\begin{eqnarray}
y^\prime=\Tilde{y}^\prime = 0 \, ,\; \;
y_{1A}=y_{2B}=\Tilde{y}_{1A}=\Tilde{y}_{2B}=0 \nonumber\\
y_{2A} = \Tilde{y}_{2A} \equiv y \, ,\; \;
y_{1B}=\Tilde{y}_{1B} \equiv \Tilde{y} \, ,\; \; y \gg \Tilde{y}
\label{eq:YukAnsatz}
\end{eqnarray}
This is equivalent to choosing the Yukawa matrix to be diagonal in a certain basis. There are two Dirac fermions which we call $N_h$ with mass $y\langle \Phi \rangle$, and two much lighter Dirac fermions with mass $\tilde{y}\langle \Phi \rangle$ which we call $N_l$. The $N_l$ 
will have masses comparable to the reheating temperature after inflation. The $N_h$, which couple much more strongly to the inflaton, will have important loop contributions to the inflaton potential.




As this is a Coleman Weinberg-type model, we also suppress the bare mass term for the Higgs. This mass is then generated dynamically by the $\lambda_{\Phi H}$ operator in the Lagrangian. In order to get the correct Higgs mass, we require
\begin{equation}
    \lambda_{\Phi H} \sim -\frac{m_H^2}{\langle \Phi \rangle^2} \,
    \label{eq:phihvalue}
\end{equation}
which, as we will see, means that we will have $\lambda_{\Phi H} \sim 10^{-17}$. As such, it will play no role in the phenomenology besides dynamical generation of the Higgs mass parameter. This value is radiatively stable: if $\lambda_{\Phi H} \rightarrow 0$ then the SM and the inflation sector are decoupled. Therefore the running of $\lambda_{\Phi H}$ under renormalization group (RG) must be proportional to $\lambda_{\Phi H}$.  The leading term in the $\beta$ function goes like $g_2^2 \lambda_{\Phi H}$
where $g_2$ is the $SU(2)$ gauge coupling in the SM. Approximating the RG by assuming $g_2$ is a constant and all other terms can be neglected, we get
\begin{equation}
  \lambda_{\Phi H} (\mu) \sim
  \lambda_{\Phi H} (\mu')
  \left(1+\frac{g_2^2}{16 \pi^2}\log(\mu/\mu')\right) \, .
\end{equation}
This means that $\lambda_{\Phi H}$ only varies by a few percent as we run from the inflation scale to the scale of $\langle \Phi \rangle$. 

At tree-level, the inflaton potential is simply $\lambda_0 \Phi^4$, but of course the loop corrections are necessary. At one loop, the $\beta$-function for the quartic is given by:
\begin{align}
     \beta_{\lambda_\Phi} &= \kappa \left(\frac{9}{8}g^4 - 4 y^4 - 4 \Tilde{y}^4 - 
     (9g^2 - 8y^2 - 8\Tilde{y}^2)\lambda_\Phi + 2\lambda^2_{\Phi H} + 24\lambda^2_{\Phi}\right), 
    \label{eq:beta}
\end{align}
where $\kappa = (16\pi^2)^{-1}$, $g$ is the $SU(2)_D$ gauge coupling, and $y$ and $\tilde{y}$ are the Yukawa couplings (see Eq.~\eqref{eq:YukAnsatz}), and we have taken $\tilde{y} \ll y$. 
For $\beta_{\lambda_\Phi}$ to have zeros, we need $y \simeq g$ . Furthermore, in order to get the correct amplitude for the matter power spectrum, $\lambda_\Phi$ must be very small, and we therefore use the parametric regime $y \simeq g \gg \lambda_\Phi$.  From Eq.~\eqref{eq:phihvalue}, we also have that 
$\lambda_{\Phi H} \lesssim \lambda_\Phi$. 
Therefore, we can neglect the terms in $\beta_{\lambda_\Phi}$ that depend on either of the scalar quartics (and we can also ignore terms proportional to $\tilde{y}$). The running of $\lambda_\Phi$ is thus controlled by the running of $g$ and $y$ whose $\beta$-functions are given by
\begin{widetext}
\begin{align}
     \beta_g &= -\kappa \left(\frac{43}{6} - \frac{2}{3}n_f\right)g^3,\nonumber \\
     \beta_y &= \kappa \left( \frac{7}{2} y^3 + \frac{1}{2} y\Tilde{y}^2 - \frac{9}{4}g^2 y\right),  
     \label{eq:betagy}
\end{align}
\end{widetext}
where $n_f$ is the number of Dirac fermion doublets kinematically accessible at a given scale.\footnote{We include the $\tilde{y}$ dependence in these equations for completeness, but can ignore its effects.} We construct a model with two separate zeros for $\beta_{\lambda_{\Phi}}$ with a massive threshold of fermions~\cite{Bai:2020zil} such that above the threshold, the gauge coupling has $\beta_g>0$, while below the threshold the theory is asymptotically free, $\beta_g<0$. If at the threshold $g\lesssim y$ and $g$ runs faster than $y$ both above and below the threshold, then this will achieve the desired behaviour for $\lambda_\Phi$ as a function of energy scale.
In the Lagrangian given in Eq.~\eqref{eq:Lag1}, the  threshold is given by $M$. As shown in Table~\ref{tab:model}, we choose $n_f=1$ below the scale $M$, and $n_f = N_\psi + 1$ above the scale, with the requirement that $N_\psi \geq 10$ so that $\beta_g>0$ above the scale $M$.

We can now integrate the one-loop $\beta$-functions to get leading-log solutions to the running couplings in the limit where terms containing $\lambda_{\Phi}$ and $\lambda_{\Phi H}$ in Eq.~\eqref{eq:beta} can be ignored. If we define our reference scale $\phi_0$ as the scale where $\beta_{\lambda_\Phi}=0$, then at that scale we have\footnote{This is the one-loop condition for the inflection point and will get modified at higher order. For example at two loops we have 
\begin{equation}
    y_0 = \frac{\sqrt{3}}{2^{5/4}}g_0 \left( 1+ \frac{620+33\sqrt{2}}{96} \kappa g_0^2\right) .
\end{equation}
Since $\kappa g_0^2 \ll 1$, the one loop expression is a good approximation. 
} 
\begin{equation}
y_0=\frac{\sqrt{3}}{2^{5/4}} g_0 \, ,
\label{eq:betazero}
\end{equation}
where the $0$ subscripts means those couplings evaluated at the scale $\phi_0$. As discussed above, the condition in Eq.~\eqref{eq:betazero} is satisfied at two different points in field space, so we take our boundary for RG running to be the lower one with $\phi_0<M$. The solution for the quartic coupling at leading log is then given by
\begin{equation}
\lambda_{\Phi}(\Phi) = \lambda_0+\kappa^2 g_0^6 \left(-b_\lambda \ln^2\frac{\Phi}{\phi_0} + \Theta(\Phi-M) \frac{3}{2}N_\psi \ln^2 \frac{\Phi}{M} \right) \, ,
\end{equation}
where $\Theta(x)$ is the Heaviside step function, $g_0$ and $\lambda_0$ are the values of those couplings at the scale $\phi_0$, and we have used Eq.~\eqref{eq:betazero} to eliminate the boundary value of $y$. Here $b_\lambda = 9 (68+21\sqrt{2})/64 \approx 13.7$.
The above solution is a good approximation if 
$\lambda_\Phi, \lambda_{\Phi H} \ll g^2,y^2$, 
which is satisfied in our phenomenologically viable parameter space as we will see. We can now plug $\lambda_\Phi$ into our potential in Eq.~\eqref{eq:schematic}. We follow~\cite{Bai:2020zil} and use the following phenomenological form of the potential:
\begin{equation}
    V(\Phi) = -\frac{a}{4}\Phi^4\left[1+b\ln^2\left(\frac{\Phi}{\phi_0}\right) \right. 
    - \left.c\ln^2\left(\frac{\Phi}{M}\right)\Theta(\Phi-M)\right] + aV_0 \; .
    \label{eq:potp}
\end{equation}
We have added a cosmological constant term $V_0$ in order for the minimum of the potential to be at (nearly) zero energy. As long as $a>0$, $c > b>0$, and $M>\phi_0$, this potential will have a broad plateau around $\phi \sim \phi_0$ and then a stable minimum with $\langle \Phi \rangle > M > \phi_0$. The phenomenological parameters are controlled by the field theory parameters, $a$ by $\lambda_0$, $b$ by $g_0$, and $c$ by $N_\psi$. We note that the Heaviside function in the potential is a direct consequence of the heavy threshold at the scale $M$.

\begin{figure}[!h]
    \centering
\includegraphics[height=7 cm, width=8 cm,angle=0]{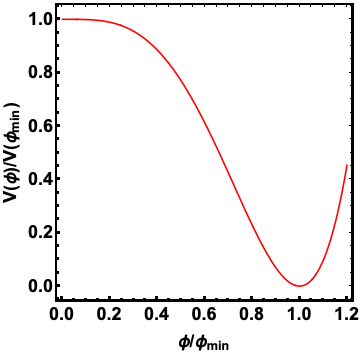}
\includegraphics[height=7 cm, width=8 cm,angle=0]{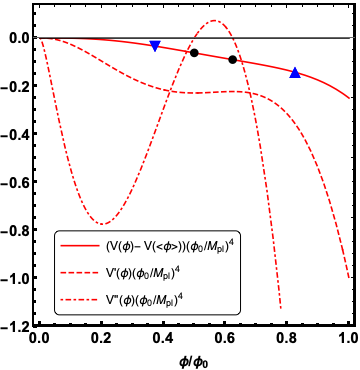}
    
    \caption{The inflaton potential. The left panel shows the full potential, while the plot on the right is a close up of the region near the inflection points. The plots are scaled such that they look the same for all allowed parameters. The right panel also shows the 1st (2nd) derivative of the potential dashed (dot-dashed), the location of the inflection points (black dots), and the points where cosmological scales enter the horizon (triangles) for the different benchmark points. The triangle on the left (right) is for BP1 (BP2) with negative (positive) $\alpha$ (see Table~\ref{tab:inflation}).}
    \label{fig:potential}
\end{figure}

In order to have the successful inflation scenario there are two additional restrictions on $b$. First, the potential cannot not develop any local minima to ensure that the inflation rolls smoothly towards the global minima: $V^\prime(\phi) < 0$ for $\phi< \langle \Phi \rangle$. This imposes $b<16$. Second, at field values $\phi\sim \phi_0 < M$, the potential should have inflection points, $V^{\prime \prime}=0$, which means that $b\geq 144/25\approx 5.7$. The inflaton potential is shown in Fig.~\ref{fig:potential} with a close up of the region where cosmological scales enter the horizon in the right panel. We can then find the parameters of the inflaton potential, including the point on the potential where cosmological scales enter the horizon, $\phi_i$, such that 60 e-folds can be achieved and the observations of $n_s$ and $A_s$ can be matched (see appendix~\ref{app:review} for more details). As described in~\cite{Bai:2020zil}, there are two solutions, one at lower field values than the inflection points and one at higher field values. These in turn correspond to different signs of the running of the spectral index $\alpha$ and is shown in the right panel of Fig.~\ref{fig:potential}.  Two detailed benchmark are explored further in the following sections and detailed in Table~\ref{tab:inflation}.

The field content presented in this section, in addition to giving an attractive inflation model that is consistent with observations from the CMB, can also solve several of the problems associated with the Standard Model as we now describe in the following sections.

\section{Neutrino Mass}
\label{sec:neutrino}

The dynamical inflection point inflation requires the existence of fermions that are singlets under the dark and the SM gauge symmetry that we have denoted $\chi$. These fermions can couple to the SM leptons the same way that right handed neutrinos do in many models. Unlike in typical seesaw models, the $\chi$ fields also get a Dirac mass with the $\psi_1$ fields via the vev of the inflaton. Ultimately these fields will still be responsible for giving the neutrinos mass. This neutrino mass generation resembles the inverse seesaw mechanism~\cite{Mohapatra:1986aw, Mohapatra:1986bd, Gonzalez-Garcia:1988okv}. The right-handed-neutrino-like states will also participate in cogenesis of the baryon asymmetry of the universe  and dark matter as described in detail in section~\ref{sec:lepto}. 


In order to generate light neutrino masses, we consider mass and Yukawa terms involving heavy fermions $\chi, \psi_1$ that  break the $Z_3$ global symmetry. In order not to disturb the inflationary dynamics, the bare mass terms of these fermions are required to be smaller than inflaton field values at the start of inflation. This requires such bare mass terms to be $\leq \phi_i \sim \mathcal{O}(\rm \textcolor{blue}{GeV})$. Since the bare mass terms are required to be small, we consider the inverse seesaw realisation where the bare mass term arises only for singlet fermion $\psi_1$ which does not couple directly to SM lepton doublets. 
The operators required are written as follows:
%
%
\begin{widetext}
\begin{align}
-\mathcal{L}_\nu & = 
(Y_R)_{\alpha i} \overline{L}_\alpha \tilde{H} \chi_{iR} + 
(Y_L)_{\alpha i} \overline{L}_\alpha \tilde{H} \chi^c_{iL} + 
 \frac{1}{2} \mu \,\overline{\psi_{1L}} \psi_{1R} +{\rm h.c.}
\label{eq:nuYukawa}
\end{align}
\end{widetext}
Here $\alpha$ is a lepton flavour index while $i$ is a $\chi$ flavour index. 

We can analyze the fermion mass matrix after the inflaton and Higgs settle to their non-zero vacuum values. Using the ansatz of Eq.~\eqref{eq:YukAnsatz}, the heaviest fermions are the $\psi_2$ and $\chi_2$, which get a Dirac mass of $y\langle \Phi \rangle$ and can be integrated out. 
We can then write the mass matrix for the remaining 
fermions in the  basis $\{ \nu_\alpha,\chi_{1R},\chi_{1L},\psi_{1L_1},\psi_{1R_1}\}$ as follows:
\begin{equation}
M_f = \begin{pmatrix}
0 &  Y\langle H \rangle  & Y\langle H \rangle & 0 & 0 \\
Y\langle H \rangle & 0 & 0 & \Tilde{y} \langle \Phi \rangle & 0\\
Y\langle H \rangle &  0  & 0 & 0 & \Tilde{y} \langle \Phi \rangle \\
  0 &  \Tilde{y} \langle \Phi \rangle  & 0 & 0 & \mu \\
  0 &  0  & \Tilde{y} \langle \Phi \rangle & \mu & 0 \\
\end{pmatrix}.
\label{eq:numatrix}
\end{equation}
We have used $Y \sim Y_{R,L}$ to simplify the notation of the Yukawa couplings to the SM neutrinos. 
Since $\langle \Phi \rangle \gg \langle H \rangle, \mu$, it is natural to consider the hierarchy $ \tilde{y} \langle \Phi \rangle \gg Y \langle H \rangle$.  
This mass matrix has the structure of the inverse seesaw mechanism, and we can thus give the neutrino mass 
%
\begin{equation}
m_\nu \approx 2\mu 
\left( \frac{Y \langle H \rangle}{\tilde{y} \langle \Phi \rangle}
\right)^2 \, .
\label{eq:numass}
\end{equation}
The heavy mass scale $\tilde{y} \langle \Phi \rangle$ must be comparable to the reheating temperature for successful leptogenesis as well as production of dark matter (see section~\ref{sec:lepto}). Therefore it will be $\sim$ TeV. If the neutrino Yukawa couplings are small, then, in order to get the observed neutrino masses, we need larger soft mass $\mu$ and vice versa. In our benchmarks we will use 
$Y\langle H \rangle \sim 10^{-3}$ GeV 
and thus require
$\mu \sim 0.1$ GeV, 
which is small enough not to affect the inflationary dynamics. The $\nu$ states will have a small mixing with the $\psi_1$ states which is of order $\sim Y\langle H \rangle / (\tilde{y}\langle \Phi \rangle)$.

Even with the simplified parameterization we have chosen, all the neutrino flavour structure can be encoded in $Y$, the Yukawa couplings of the SM neutrinos to the $\chi$'s. As there are two $\chi$ states after integrating out those at $y \langle\Phi \rangle$, there will still be three different mass eigenvalues. We leave a complete analysis of the neutrino flavour sector to future work.

\section{Cogenesis of lepton asymmetry and dark matter}
\label{sec:lepto}

The ingredients of the inflation model are also sufficient for the cogenesis of both a baryon asymmetry and dark matter. The right handed neutrinos that give the neutrinos mass as described in the previous section are also the key players for the cogenesis. The baryon asymmetry will be seeded via the leptogenesis mechanism~\cite{Fukugita:1986hr} with asymmetric decays of the lightest right handed neutrino generating a lepton asymmetry. The baryon asymmetry will then be formed by the usual electroweak sphaleron process~\cite{Kuzmin:1985mm}. The dark matter production will proceed via freeze-in with two processes contributing to its abundance: decay of the inflaton and
annihilation of the right handed neutrinos. We will now describe these processes in detail and solve the coupled Boltzmann equations for all the relevant states (including the inflaton) to show that the correct BAU and dark matter abundance can be achieved.

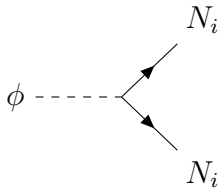
\begin{figure*}[!h]
    \centering
    \begin{tabular}{lr}
    \begin{tikzpicture}[/tikzfeynman/small]
      \begin{feynman}
      \vertex (v1){$\phi$};
      \vertex[right = 1.41cm of v1](v2);
      \vertex[above right = 1.cm of v2](i){$N_i$};
      \vertex[below right = 1.cm of v2](j){$N_i$};
      \diagram*[small]{(v1)--[scalar](v2)--[fermion](i),(v2)--[fermion](j)};
      \end{feynman}   
      \end{tikzpicture}
      \end{tabular}
    \caption{Process responsible for reheating.}
    \label{fig:reheat}
\end{figure*}

As noted previously, there are four approximately degenerate fermions with mass around the reheating temperature $\sim$ TeV that are made of $\psi_1$ and $\chi$. They get a large Dirac mass given by $\tilde{y} \langle \Phi \rangle$ (see Eqs.~\eqref{eq:Lag1} and \eqref{eq:YukAnsatz}) and also get a small splitting due to the symmetry breaking term $\mu$ (see Eq.~\eqref{eq:nuYukawa}).  We denote these states $N_i$ with $N_1$ being the lightest. After inflation, the universe will be reheated by inflaton decays. The main decay mode is $\phi \rightarrow N_iN_i$ shown in Fig.~\ref{fig:reheat}. 
Other decay modes include $\Phi \rightarrow HH$ and $\Phi \rightarrow \bar{\psi}_i\psi_i$. Both are subdominant in the total width, but, as we will see below, the latter decay will be an important contribution to the dark matter density.
After the inflaton decays, the universe will be reheated into a thermal bath that contains $N_i$ states and all states with large couplings to the $N$'s. This will include all of the SM states via the Yukawa coupling in Eq.~\eqref{eq:nuYukawa}. The reheating temperature will be $\sim$ TeV and in the instantaneous reheating approximation, it is set by the inflaton width (see Eq.~\eqref{eq:reheat}). We do not use that approximation and instead calculate the reheating temperature explicitly by solving the relevant Boltzmann equations for the inflaton, the radiation bath, the lightest right handed neutrino, DM, and the $B-L$ asymmetry. The 
reheating temperature is calculated by finding the time where radiation energy density equals that of the inflaton. After that time, radiation dominates the universe and its temperature redshifts as $T \propto 1/a$.

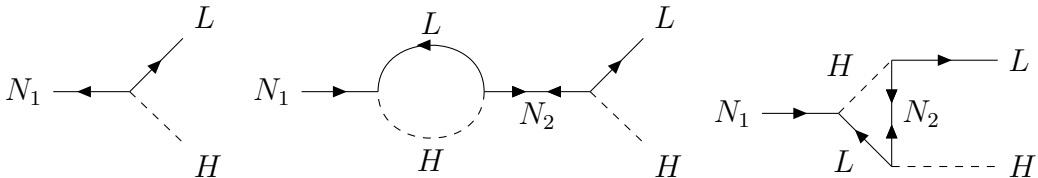
\begin{figure*}[!h]
    \centering
    \begin{tabular}{lcr}
    \begin{tikzpicture}[/tikzfeynman/small]
      \begin{feynman}
      \vertex (v1){$N_1$};
      \vertex[right = 1.41cm of v1](v2);
      \vertex[above right = 1.cm of v2](i){$L$};
      \vertex[below right = 1.cm of v2](j){$H$};
      \diagram*[small]{(v1)--[anti fermion](v2)--[fermion](i),(v2)--[scalar](j)};
      \end{feynman}   
      \end{tikzpicture}    
         &  
    \begin{tikzpicture}[/tikzfeynman/small]
      \begin{feynman}
      \vertex (v1){$N_1$};
      \vertex[right = 1.41cm of v1](v2);
      \vertex[right = 1.41cm of v2](v3);
      \vertex[right = 1.41cm of v3](v4);
      \vertex[above right = 1.cm of v4](i){$L$};
      \vertex[below right = 1.cm of v4](j){$H$};
      \diagram*[small]{(v1)--[fermion](v2)--[anti fermion,half left,looseness = 1.5,edge label = $L$](v3)--[scalar,half left,looseness = 1.5,edge label = $H$](v2),(v3)--[majorana,edge label' = $N_2$](v4),(v4)--[fermion](i),(v4)--[scalar](j)};
      \end{feynman}   
      \end{tikzpicture}    
          &  
    \begin{tikzpicture}[/tikzfeynman/small]
      \begin{feynman}
      \vertex (v1){$N_1$};
      \vertex[right = 1.41cm of v1](v2);
      \vertex[above right = 1.cm of v2](v3);
      \vertex[below right = 1.cm of v2](v4);
      \vertex[right = 1.41cm of v3](j){$L$};
      \vertex[right = 1.41cm of v4](k){$H$};
      \diagram*[small]{(v1)--[fermion](v2)--[anti fermion,edge label' = $L$](v4)--[majorana,edge label' = $N_2$](v3)--[fermion](j),(v2)--[scalar,edge label = $H$](v3),(v4)--[scalar](k)};
      \end{feynman}   
      \end{tikzpicture}    
    \end{tabular}
    \caption{Process responsible for leptogenesis.}
    \label{fig:lepto}
\end{figure*}

 The $N_1$ decay is dominated by the process $N_1 \rightarrow LH$ as shown in Fig.~\ref{fig:lepto}. The decay occurs out of equilibrium and will lead to a non-zero CP asymmetry from the interference of the tree and one-loop diagrams shown in Fig.~\ref{fig:lepto}.  
Primarily the asymmetry will be coming from the resonance \cite{Pilaftsis:2003gt} which is from second diagram in Fig.~\ref{fig:lepto}. Earlier work on leptogenesis in inverse seesaw type scenarios can be found in \cite{Blanchet:2010kw, Gu:2010xc}. The CP asymmetry parameter corresponding to the CP violating decay of RHN $N_i$ (summing over all lepton flavours) is given by \cite{Pilaftsis:2003gt}
\begin{eqnarray}
\epsilon_{i} & \equiv & \dfrac{\Gamma_{(N_{i}\longrightarrow \sum_{\alpha} L_{\alpha} H )}-\Gamma_{(N_{i}\longrightarrow \sum_{\alpha} L_{\alpha}^{c}H^\dagger)}}{\Gamma_{(N_{i}\longrightarrow \sum_{\alpha}L_{\alpha}H)}+\Gamma_{(N_{i}\longrightarrow \sum_{i}L_{\alpha}^{c}H^\dagger)}}  \\ 
 & = & \sum_{j \neq i, j=1}^4 \dfrac{{\rm Im}[(h^{\dagger}h)_{ij}^{2}]}{(h^{\dagger}h)_{ii}(h^{\dagger}h)_{jj}}\dfrac{(M_{N_i}^{2}-M_{N_j}^{2})M_{N_i}\Gamma_{N_j}}{(M_{N_i}^{2}-M_{N_j}^{2})^{2}+M_{N_i}^{2}\Gamma_{N_j}^{2}}. \label{eq:asymmparameter} 
\end{eqnarray} 
Here we denote the Yukawa coupling of lepton flavour $\alpha$ to $N_i$ as $h_{\alpha i}$ which can be found from the ones in Eq.~\eqref{eq:nuYukawa} by going to the physical mass basis of heavy fermions. Now, parametrizing the Yukawa couplings as $h_{\alpha i} = y_{\alpha i} e^{-i\theta_i}$ and considering the resonant limit
$\Delta M_{ij} \equiv M_{N_i}-M_{N_j} \sim \Gamma_{N_j}$ we can approximate the CP asymmetry parameter as 
\begin{equation}
    \epsilon_i \sim \frac{3}{2}\sin(\theta), \quad \theta_i - \theta_j =\theta.
\end{equation}
If $\Delta M$ and $\Gamma$ are not comparable, then there is additional suppression by $\sim \text{Min}(\Delta M / \Gamma,\, \Gamma/\Delta M)$.
Our benchmarks described below have large values of $\epsilon$ and require the parameters to be in the resonance regime. There are also regions of parameter space where the model is viable with $\epsilon \ll 1$ where there is freedom in the interplay of $\Delta M$, $\Gamma$ and $\theta$.

When the lepton asymmetry is generated, some of the asymmetry will be converted to a baryon asymmetry via  the electroweak sphaleron process~\cite{Kuzmin:1985mm}. This process conserves $B-L$, and it is fast for temperatures above $T \sim 130$ GeV, but exponentially suppressed for lower temperatures. The equilibrium value of the baryon asymmetry is related to the $B-L$ asymmetry by the sphaleron conversion factor given as~\cite{Harvey:1990qw}
\begin{align}
& a_\text{sph}=\frac{8\,N_F+4\,N_H}{22\,N_F+13\,N_H}\,,
\label{eq:sphaleron}
\end{align}
where $N_F$ the number of fermion generations and $N_H$ is the number of Higgs doublets that transform under the $SU(2)_L$ gauge symmetry of the SM. Since we do not have any additional $SU(2)_L$ multiplets in our model, we have $N_F=3\,,N_H=1$ and $a_\text{sph}= 28/79$.

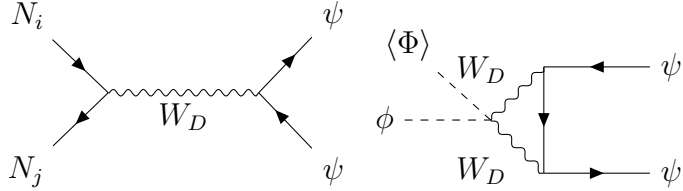
\begin{figure}[!h]
    \centering
    \begin{tabular}{lr}
      \begin{tikzpicture}[/tikzfeynman/small]
      \begin{feynman}
      \vertex (v1);
      \vertex[above left = 1. cm of v1](i){$N_i$};
      \vertex[below left = 1. cm of v1](j){$N_j$};
      \vertex[right= 2.cm of v1](v2);
      \vertex[above right= 1.cm of v2](k){$\psi$};
      \vertex[below right = 1.cm of v2](l){$\psi$};
      \diagram*[small]{(i)--[fermion](v1)--[fermion](j),(v1)--[photon,edge label'=$W_D$](v2),(k)--[anti fermion](v2)--[anti fermion](l)};
      \end{feynman}   
      \end{tikzpicture} &
      \begin{tikzpicture}[/tikzfeynman/small]
      \begin{feynman}
      \vertex (v1){$\phi$};
      \vertex[right = 1.41cm of v1](v2);
      \vertex[above left = 0.9cm of v2](l){$\langle \Phi \rangle$};
      \vertex[above right = 1.cm of v2](v3);
      \vertex[below right = 1.cm of v2](v4);
      \vertex[right = 1.41cm of v3](j){$\psi$};
      \vertex[right = 1.41cm of v4](k){$\psi$};
      \diagram*[small]{(l)--[scalar](v2),(v1)--[scalar](v2)--[boson,edge label' = $W_D$](v4),(j)--[fermion](v3)--[fermion](v4)--[fermion](k),(v2)--[boson,edge label = $W_D$](v3)};
      \end{feynman}   
      \end{tikzpicture} 
      \end{tabular}
    \caption{Processes responsible for dark matter freeze-in. }
    \label{fig:dm}
\end{figure}

As noted above, there are two sources of dark matter production. 
The first is via annihilation of the $N_i$ states as shown by the tree level Feynman diagram on the left panel of Fig. \ref{fig:dm}. The second is from inflaton decay.
While the inflaton does not couple directly to dark matter, such a decay can arise at one-loop level as shown on the right panel of Fig. \ref{fig:dm}. The dark matter are the $\psi_i$ shown in Table~\ref{tab:model}. There are $N_\psi$ generations that are stable due to being the only states charged under the unbroken $Z_2$ symmetry. 
These states, however do not couple to the SM directly nor have any tree level production (or annihilation) channels from (to) the SM bath particles. The DM does have $SU(2)_D$ gauge interactions, but the corresponding gauge bosons have mass $g \langle \Phi \rangle$ which is much larger than $T_\text{reheat}$, and are therefore not present in the bath after reheating. Since the inflaton condensate scale is very large compared to the reheating scale, even its loop suppressed decay to DM can contribute significantly to the abundance of the latter. As we will see below, the two processes have very different parametric dependence and each can be important in various regions of parameter space. Both processes are slow and do not reach equilibrium, so this model is in the feebly interacting massive particle (FIMP) paradigm \cite{Hall:2009bx}.

%

In order to study the generation of lepton asymmetry and dark matter relic abundance, one has to write the corresponding Boltzmann equations. We begin with a universe dominated by the inflaton at the end of inflation. At this stage the inflaton red-shifts like matter, and its decay sources the rest of the thermal bath. We also track the abundances of the $N_i$ and $\psi_i$, as well as  the asymmetry in $B-L$ number so that sphaleron dynamics can be ignored. The total baryon asymmetry can then be found by using the sphaleron factor from Eq.~\eqref{eq:sphaleron}.

We write the coupled Boltzmann equations for energy density of the inflaton condensate $\rho_\phi$ during the oscillatory phase, energy density of the lightest RHN $\rho_N$, number density of $B-L$ $n_{B-L}$, radiation density $\rho_R$, DM energy density $\rho_\psi$ as \cite{Giudice:2000ex, Hahn-Woernle:2008tsk, Garcia:2020eof, Barman:2021ost}:
\begin{align}
& \dot{\rho_\phi} = -3\,\mathcal{\mathcal{H}}\,\rho_\phi-\Gamma_\phi\,\rho_\phi,\\ &
\dot{\rho_N} = -(3 + \Theta(3T-M_{N_1}))\mathcal\,{\mathcal{H}}\,\rho_N+ \Gamma_{\phi \rightarrow NN} \,\rho_\phi-\Gamma_N\, \rho_N, \\ &
\dot{n}_\text{B-L} = -3\,\mathcal{\mathcal{H}}\,n_\text{B-L}+\epsilon_1 \,\frac{\Gamma_N}{\langle E_N \rangle}\,(\rho_N-\rho^{\rm eq}_N)-\Gamma_N\frac{n^{\rm eq}_N}{n^{\rm eq}_\gamma}\,n_\text{B-L},\\ &
\dot{\rho}_R = -4\mathcal{H}\,\rho_R+\Gamma_N\, \rho_N,   
\\   &
\dot{\rho_\psi} = -(3 + \Theta(3T-M_\psi))\mathcal\,{\mathcal{H}}\,\rho_\psi+ 2\frac{\langle \sigma v \rangle}{\langle E_N \rangle} \,\rho^2_N + \Gamma_{\phi \rightarrow \psi \psi} \rho_\phi.
\end{align}
We consider the inflaton condensate to redshift as matter which is valid during the oscillatory phase after inflation ends. For the lightest RHN $N$ and the DM $\psi$, we use a Heaviside $\Theta$ function to model the transition from radiation to matter at a temperature equal to one third the mass~\cite{Easa:2022vcw}. 
In order to solve these coupled equations, we re-scale the energy and number densities with scale factor $a$ as
\begin{align}
    \widetilde{\rho}_\phi = \rho_\phi a^3, \; \widetilde{\rho}_N = \rho_N a^3, \; \widetilde{N}_{B-L} = n_{B-L} a^3, \; \widetilde{\rho}_\psi = \rho_\psi a^3, \; \widetilde{\rho}_R = \rho_R a^4. 
\end{align}
With the initial scale factor as $a_I$, we define $\xi=\frac{a}{a_I}$ as the ratio of the scale factors leading to the expression of the Hubble parameter as
\begin{align}
    \mathcal{H} &= \sqrt{\frac{8\pi}{3M^2_{pl}}\frac{\widetilde{\rho}_\phi a_I \xi + \widetilde{\rho}_N a_I \xi + \widetilde{\rho}_R}{a^4_I\xi^4}}, \quad \left.\widetilde{\rho}_\phi\right|_{(\xi=1)} =  \rho_0 \simeq \frac{1}{4}\lambda_0 \langle \Phi \rangle^4 .
\end{align}
We have ignored the contribution of the dark matter to Hubble, which is a good approximation in the freeze-in regime. We assume $a_I=1$ for simplicity, though the final result does not depend upon this choice. 

The quantity $\langle E_f \rangle$ that appears in the Boltzmann equations is  average energy of a species $f$ and is defined as
\begin{align}
    \langle E_f \rangle = \rho^{\rm eq}_f/n^{\rm eq}_f, \;\; (\rho^{\rm eq}_f,n^{\rm eq}_f) &= g_f\int^\infty_0\frac{d^3p}{(2\pi)^3} \frac{(E,1)}{e^{E/T} \pm 1 }.
\end{align}
As the inflaton condensate's energy is converted into radiation, the temperature of the universe can be computed from $ T^{-1} = \left(\frac{\pi^2 g_*(T)}{30\rho_R}\right)^{1/4}$. In terms of re-scaled densities, we can re-write the above equations as
\begin{align}
    \widetilde{\rho}^\prime_\phi &= -\frac{1}{\mathcal{H}\xi} \Gamma_\phi \,\widetilde{\rho}_\phi,\\
    \widetilde{\rho}^\prime_N &= -\frac{\theta(3T-M_{N_1})}{\xi} \widetilde{\rho}_N +\frac{1}{\mathcal{H}\xi}\left(\Gamma_{\phi \rightarrow NN} \,\widetilde{\rho}_\phi-\Gamma_N\, (\widetilde{\rho}_N-\widetilde{\rho}^{\rm eq}_N) \right),\\
    \widetilde{N}^\prime_{B-L} &= \frac{1}{\mathcal{H}\xi}\left(\epsilon_1 \,\frac{\Gamma_N}{\langle E_N \rangle}\,(\widetilde{\rho}_N-\widetilde{\rho}^{\rm eq}_N)-\Gamma_\text{ID}\,\widetilde{N}_\text{B-L} \right),\\
    \widetilde{\rho}^\prime_R &= \frac{\Gamma_N}{\mathcal{H}}\, (\widetilde{\rho}_N-\widetilde{\rho}^{\rm eq}_N), \\
    \widetilde{\rho}^\prime_\psi &= -\frac{\theta(3T-m_\psi)}{\xi} \widetilde{\rho}_\psi + \frac{1}{\mathcal{H}\xi}\left(2\frac{\langle \sigma v \rangle}{\langle E_N \rangle \xi^3} \,\widetilde{\rho}^2_N + \Gamma_{\phi \rightarrow \psi \psi} \widetilde{\rho}_\phi \right),\label{eq:BoltzrhoPsi}
\end{align}
where $X'$ denotes the derivative of $X$ with respect to the scale factor $\xi$.

The thermally averaged cross section~\cite{Gondolo:1990dk} $\langle \sigma v \rangle$ for the annihilation process $N_i N_j \rightarrow \psi \psi$ is given by 
\begin{align}
  \langle \sigma v \rangle &= \frac{T}{128\pi^5(n^{\rm eq}_{N_1})^2}\int^\infty_{\hat{s}_0} d\hat{s} \,  p_{\psi} \, p_{N_1} \frac{|\mathcal{M}|^2}{\sqrt{\hat{s}}}K_1 \left( \sqrt{\hat{s}}/T \right), \quad \hat{s}_0 \equiv {\rm Max}\left[4M^2_{N_1},4m^2_{\psi}\right], \nonumber \\
  p_i &= \frac{\lambda^{1/2}(\hat{s},m^2_i,m^2_i)}{2\sqrt{\hat{s}}}, \; \lambda(x,y,z) = x^{2}+y^{2}+z^{2}-2xy-2yz-2xz ,\nonumber \\
 \sum |\mathcal{M}|^2 &\simeq \frac{10 N_\psi}{3\langle \Phi \rangle^4}\left(2\hat{s}^2-\frac{8}{5}\hat{s}\left(M^2_{N_1} - \frac{m^2_\psi}{4}\right) + 4m^2_\psi M^2_{N_1} \right). 
\end{align}
The one-loop decay width of the inflaton into dark matter is 
\begin{align}
    \Gamma_{\phi \rightarrow \psi \psi} &= \frac{\tilde{y}^2_{\phi \psi \psi}m_\phi}{8\pi}\left[1-\frac{4m^2_\psi}{m^2_\phi}\right]^{3/2}. \\
    \widetilde{y}_{\phi \psi \psi} &\approx \frac{5 g^4_D  m_\psi \langle \Phi \rangle}{48\pi^2 m^2_{W_D}} = \frac{5 g^2_D  m_\psi}{48\pi^2 \langle \Phi \rangle}
    \label{eq:inflatonDMdecay} 
\end{align}
The full expression for the one-loop inflaton-DM effective coupling $\tilde{y}_{\phi \psi \psi}$ is 
given in appendix \ref{app:review}, with the expression here given in the large $m_{W_D}$ limit.

Before numerically solving these equations, we first give approximate analytical solutions for the $B-L$ asymmetry and the dark matter abundance. 
For the $B-L$ asymmetry, we can first compare the decay rate of the $N_i$ to the Hubble parameter. At the time of reheating, $\mathcal{H} \sim T_{\rm reheat}^2/M_{\rm pl} \sim 10^{-10}$ GeV, while $\Gamma_{N_1}\sim Y^2 M_{N_1}\sim 10^{-7}$ GeV, where we are using rough values of the parameters in Table~\ref{tab:inflation}.\footnote{We are using $M_{\rm pl} = G_N^{-1/2} \simeq 1.2\times 10^{19}$ GeV.} 
Therefore, we are in the strong wash-out regime and the inverse decay will keep the $N$ abundance close to its equilibrium value. We can thus estimate the asymmetry following~\cite{Buchmuller:2004nz}:
\begin{align}
   K &= \frac{\Gamma_{N_i}}{\mathcal{H}(T=M_{N_i})} \sim \frac{Y^2 M_{\rm pl}}{M_{N_1}}
   \nonumber\\
   \kappa(x) &= \left(1+\frac{K^2 x^6}{75} \right)^{-1}\frac{2K}{75}x^5 \approx \frac{2}{x K} \nonumber\\
   Y_{B} (T_0) \,& \textcolor{blue}{\equiv\frac{n_{B}}{s} (T_0)}= \left.\frac{3}{4}a_{\text{sph}}\epsilon_i \frac{T}{T_{\text{reheat}}}\kappa(M_{N_i}/T)\right|_{T=130 \, {\rm GeV}} \sim  10^{-10}  
   \; .  
\end{align}
The 130 GeV temperature is where the electroweak sphaleron freezes out as discussed above and $a_{\text{sph}}$ is given in Eq.~\eqref{eq:sphaleron}. The factor of $T/T_{\text{reheat}}$ accounts for the fact that $N_i$ is produced relativistically and redshifts like radiation down to the sphaleron temperature. We see that this rough estimate gives the right order of magnitude for the observed asymmetry. In our benchmarks below we have chosen relatively large values of $\epsilon_i$. One could alternatively choose smaller $Y$ and smaller $\epsilon_i$. This would mean the $N_i$ states have weaker couplings to SM leptons, but this scenario can still accommodate the BAU.


We can also make analytical estimate of DM abundance by considering one contribution at a time. Starting with the annihilation channel $(N N \rightarrow \psi \psi)$, we use the fact that the annihilation is mediated by a very heavy gauge boson so the production will be dominated by temperatures around the reheating temperature because that's when the bath of $N$'s is the hottest. In that temperature regime with $T\sim T_{\text{reheat}}>m_\psi,M_{N_1}$, we can estimate the high temperature limit of the cross section $\langle \sigma v \rangle \sim N_\psi T^2/\langle \Phi \rangle^4$. We can also use $\mathcal{H}\sim T^2$, $\rho_N \sim T^4$ and $T\sim 1/\xi$. We can then integrate the Boltzmann equation Eq.~\eqref{eq:BoltzrhoPsi}
\begin{eqnarray}
\rho_\psi(T_{\text{reheat}}) \sim 
\frac{N_\psi M_{\text{pl}} T^7_{\text{reheat}}}{\langle \Phi \rangle^4}
\end{eqnarray}
from which we can compute the yield:
\begin{eqnarray}
Y_\psi (T_0) \equiv \frac{n_\psi}{s} (T_0)\sim \left( m_\psi^3 \frac{\rho_\psi(T_{\text{reheat}})}{T^4_{\text{reheat}}} \right) \left(\frac{T_0}{m_\psi}\right)^3
\frac{1}{s_0}
\sim 
\frac{N_\psi M_{\text{pl}}T_{\text{reheat}}^3}{\langle\Phi\rangle^4}
\frac{T_0^3}{s_0} \, .
\label{eq:Yann}
\end{eqnarray}
%
%
%
We see that for $T_{\rm reheat}\sim 1$ TeV and $\langle \Phi \rangle \sim 10^{10}$ GeV, we get $Y_\psi \sim 10^{-11}$ which is the right ballpark for 10 GeV scale mass dark matter. 


For the decay process, we can again use the instantaneous decay approximation, and then at the reheating temperature the energy density of $\psi$ is just the energy density of the inflaton times its branching ratio into $\psi$. 
\begin{eqnarray}
    \rho_\psi(T_\text{reheat}) \sim \rho_\phi(T_\text{reheat}) \frac{\Gamma(\phi\rightarrow \bar{\psi}\psi)}{\Gamma(\phi\rightarrow \bar{N}N)} \sim N_\psi \left(\frac{\tilde{y}_{\phi\psi\psi}}{\tilde{y}}\right)^2
    T_\text{reheat}^4\nonumber\\
Y_\psi (T_0)\equiv \frac{n_\psi}{s} (T_0)\sim \left( m_\psi^3 \frac{\rho_\psi(T_{\text{reheat}})}{T^4_{\text{reheat}}} \right) \left(\frac{T_0}{m_\psi}\right)^3
\frac{1}{s_0}
\sim
N_\psi \left(\frac{\tilde{y}_{\phi\psi\psi}}{\tilde{y}}\right)^2 \frac{T_0^3}{s_0}
\end{eqnarray}
where $\tilde{y}_{\phi\psi\psi}$ is defined in Eq.~\eqref{eq:inflatonDMdecay} and $\tilde{y}$ is defined in Eqs.~\eqref{eq:Lag1} and~\eqref{eq:YukAnsatz}. We see that while the yield for the decay process depends on the ratio of dimensionless couplings, that for the annihilation process depends on a more complicated ratio of dimensionful parameters in Eq.~\eqref{eq:Yann}. Therefore, the dominant process will vary across the parameter space.

\begin{figure}
    \centering
    \begin{tabular}{lr}
    \includegraphics[width=0.45\textwidth]{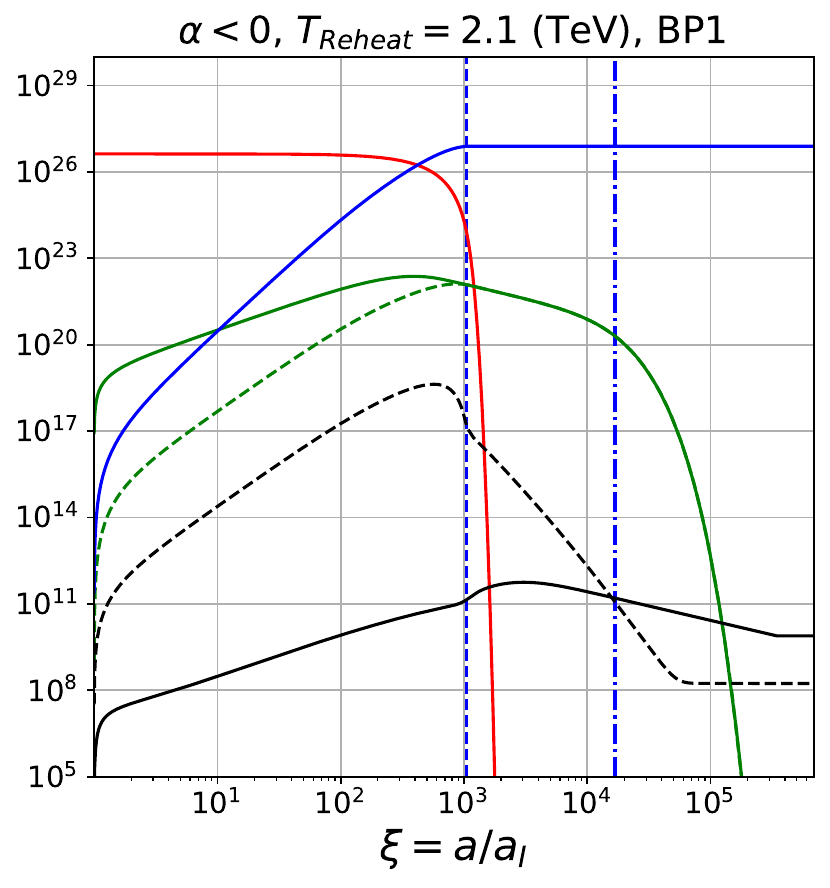}
    \includegraphics[width=0.615\textwidth]{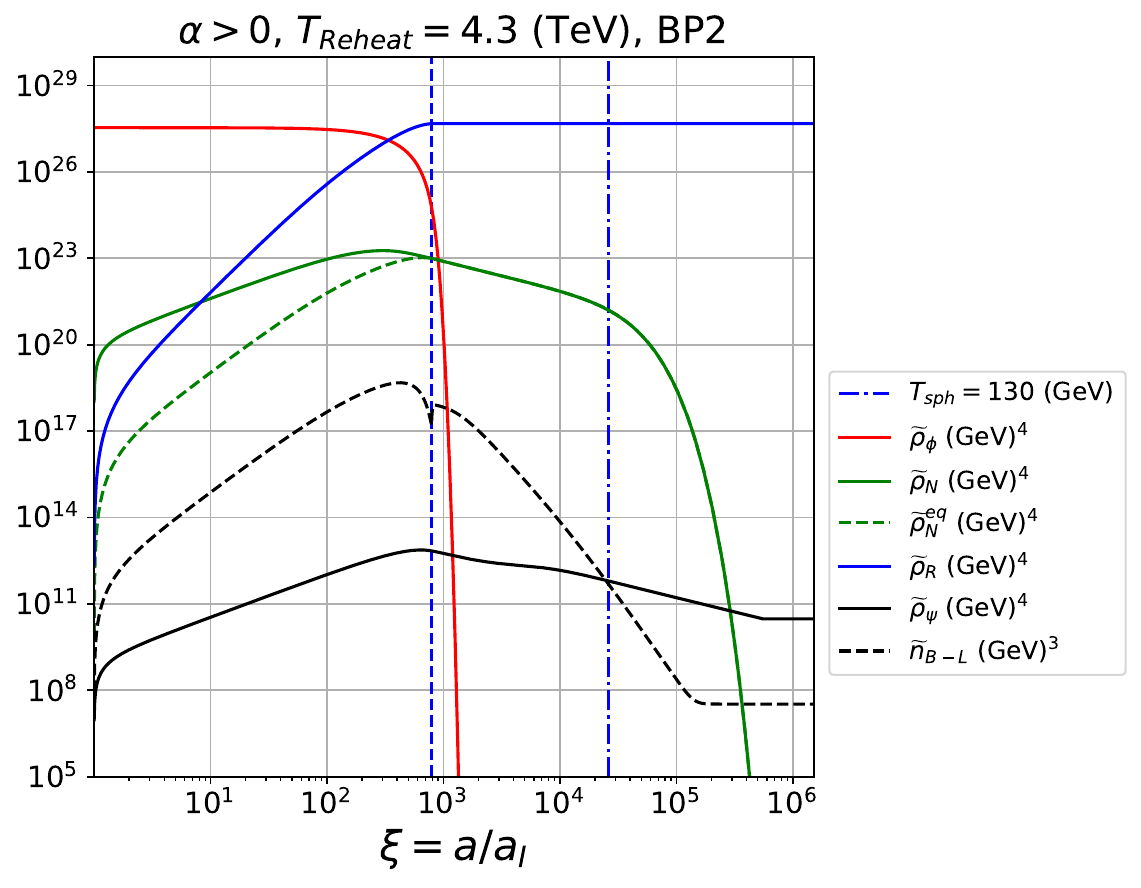}
   \end{tabular}
    \caption{Evolution of comoving energy densities for inflaton condensate $\phi$, lightest RHN, radiation, dark matter and comoving number density of $B-L$ asymmetry for the benchmark points shown in Table~\ref{tab:inflation}. The left panel corresponds to $\alpha <0$ (BP1) while the right panel corresponds to $\alpha>0$ (BP2). The vertical line on the left corresponds to the reheat temperature $T_{\rm reheat}$, when radiation energy density starts dominating and temperature redshifts as $T \propto 1/a$. The reheat temperature for each benchmark is given in the heading of each panel. The vertical line on the right corresponds to the temperature where the sphaleron decouples.} 
    \label{fig:BE_evolve2}
\end{figure}

We now do a full numerical analysis of the coupled Boltzmann equations mentioned above. We have implemented the above model using {\tt MARTY}~\cite{Uhlrich:2020ltd} and have taken the output of all the processes involved. We then used it in our own code for solving Boltzmann Equation. We explore two benchmark parameter points with all the detailed parameters and outputs shown in Table~\ref{tab:inflation}. The initial abundances of lepton asymmetry, lightest RHN, radiation as well as DM are considered to be negligible with the inflaton condensate solely contributing to the energy density of the universe. Their abundances grow as the inflaton condensate starts decaying.
The evolution of comoving energy densities for inflaton condensate $\phi$, lightest RHN, radiation, dark matter and comoving number density of $B-L$ asymmetry for different benchmark points in Table~\ref{tab:inflation} is shown in Fig.~\ref{fig:BE_evolve2}. 
One may notice that the comoving densities for radiation, DM, lightest RHN as well as $B-L$ build up from almost vanishing values as the inflaton condensate starts decaying. The reheating is complete once the radiation energy density starts dominating over the inflaton condensate and temperature redshifts with the scale factor as $T \propto 1/a$.

The figure also shows the total DM relic for $N_\psi$ copies of dark fermion doublet $\psi$ for which the final abundance matches with the Planck 2018 data \cite{Planck:2018vyg} given the masses shown in Table~\ref{tab:inflation}. We see that DM production is the largest around the epoch of reheating. The scaled energy density $\tilde{\rho}_\psi = \rho_\psi a^3$ then falls because it is produced relativistically and thus redshifts like radiation until it has cooled. For BP1, the dark matter production is dominated by the $N$ annihilation process, while for BP2, the two processes are comparable. 
Finally, we note that we have checked the reheating temperature using the instantaneous reheating approximation of Eq.~\eqref{eq:reheat} and find that it agrees with the value from the full calculation to within a few per cent for both benchmarks. 

\textcolor{blue}{\begin{table}[!ht]
    \centering
    \begin{tabular}{|c|c|c|c|c|c|c|}
      \hline    & BP1 & BP2     \\
     \hline  $M/\phi_0$   & $6.97$  &  $6.73$  \\
     \hline $\phi_0/M_{\rm pl}$ & $2.32\times 10^{-19}$ &  $2.39\times10^{-19}$ \\
     \hline $\phi_{i}$ (GeV) & $2.34$ & $1.1$ \\
     \hline $\mu$ (GeV) & $0.7$ & $0.11$ \\
     \hline $\lambda_0$ & $1.92\times10^{-14}$ & $6.92\times 10^{-13}$ \\
     \hline  $N_\psi$ & 13 & 13 \\ 
     \hline $\tilde{y} $ & $6.3\times 10^{-8}$ & $6.19\times 10^{-8}$\\
     \hline $Y$ & $1.7\times 10^{-5}$ & $2.86\times 10^{-5}$\\
     \hline $\epsilon_1$ & $0.1$ & $0.03$  \\
     \hline $\langle \Phi \rangle$ (GeV) & $1.72\times10^{10}$ & $1.2\times 10^{10}$ \\
     \hline $m_\phi$ (TeV) & 4.13 & 17.1  \\
     \hline $M=m_\psi$ (GeV) & 20 & 20 \\
     \hline $M_{N_l}$ (GeV) & $545$ &  $368$ \\
     \hline $M_{N_h}$ (GeV) & $1.8\times 10^8$ &  $2.22 \times 10^8$ \\
     \hline $T_{\text{reheat}}$ (TeV) & $2.1$ & $4.3$\\
     \hline $N_e$ & 60 & 60 \\
     \hline $n_s$ & 0.9691 & 0.9691 \\
     \hline $r$ & $4.6\times10^{-45}$ & $1.3\times10^{-44}$ \\
     \hline $\alpha$ & -$8.49\times10^{-4}$ & $2.35\times10^{-3}$ \\
     \hline
    \end{tabular}
    \caption{Details of the two benchmark points used in the analysis. 
    }
    \label{tab:inflation}
\end{table}}

\section{Conclusions}
\label{sec:conclude}

The recently proposed framework of Dynamical Inflection Point Inflation~\cite{Bai:2020zil} provides a way to generate an inflaton potential using ordinary field theory ingredients, and the inflation scale can be parametrically lower than the Planck scale and agree with all observations from the CMB. The inflaton is coupled to gauge fields and fermions, and in this work we have shown that the fields in this inflation scenario can also solve three of the most significant shortcomings of the Standard Model: neutrino masses, the baryon asymmetry of the universe, and dark matter. 

The model contains an $SU(2)_D$ gauge group, under which the inflaton is a doublet, and fermions that are singlets and doublets. All the new fields are neutral under the SM gauge symmetry. The singlet fermion can couple to the SM lepton portal $LH$. When the inflaton settles to its minimum, the singlet and some of the doublet fermions will get Dirac masses, but they have properties similar to right handed (RH) neutrinos.
These states will then give mass to the SM neutrinos via the inverse seesaw mechanism. The remaining doublets are charged under a $Z_2$ symmetry. They are thus stable and serve as the dark matter candidate.

The dominant inflaton decay is via the Yukawa couplings to the right handed neutrinos which reheats the universe up to the TeV scale.
The dynamics of the fermions also gives rise to cogenesis of a baryon asymmetry and dark matter abundance. These RH neutrinos decay asymmetrically and out of equilibrium to $LH$, generating a lepton asymmetry, which is then partly converted to a baryon asymmetry by the electroweak sphaleron. The dark sector fermions can also annihilate, via the heavy gauge boson, to the dark matter states. These states are stabilized by a $Z_2$ symmetry, but their production is rare because the gauge bosons are orders of magnitude heavier than the fermions or the dark matter. Therefore the abundance is set by the freeze-in mechanism.

Because the inflation scale is around the weak scale, we predict no discovery of tensor modes, $r \ll 1$. The expected value of the running of the scalar spectral index, $\alpha$, could be within the range of next generation CMB observations~\cite{SimonsObservatory:2018koc}. If $\alpha$ is measured, its sign would determine where on the inflaton potential cosmological scales enter the horizon (see Fig.~\ref{fig:potential}). There is a direct coupling of the inflaton to the Higgs that dynamically generates the Higgs mass parameter. If instead the portal coupling $\lambda_{\Phi H}$ is larger and there is an additional bare Higgs mass parameter, the inflaton could possibly be probed via its mixing with the Higgs~\cite{Dawson:2022zbb}. 
The RH leptons are at the weak scale and couple to SM leptons, so they could potentially be produced and discovered at colliders~\cite{Abdullahi:2022jlv}. Our benchmarks have maximized this coupling, which in turn maximizes the asymmetry parameter $\epsilon_i$. Precise measurements of decays of these right handed neutrino-like states could also shed light on whether they are in fact participating in a leptogenesis mechanism. The Yukawa couplings are expected to have $\mathcal{O}(1)$ phases, so there will be a loop-level contribution to the electric dipole moments of the charged leptons~\cite{Alarcon:2022ero}. 

The dark matter states are at the weak scale, but their coupling to the SM is suppressed by the heavy vector mass as well as by the mixing of heavy and light neutrinos. Therefore, like most freeze-in models, prospects for direct detection are quite limited. There may be signals in indirect detection, particularly if the mass spectrum of the dark matter is slightly non-minimal and there are long lived states that decay down to lighter ones.

We have studied two specific benchmark points shown in Table~\ref{tab:inflation}, and shown the evolution of the cogenesis for those benchmarks in Fig.~\ref{fig:BE_evolve2}. This gives non-trivial proof of an existence of viable parameter points that satisfy all the observational constraints of the inflation observables, BAU, and dark matter. Throughout we have taken relatively simple choices for the parameters, but a more complete exploration of the parameter space may uncover additional phenomenological signatures. 


\section*{Acknowledgements}
We are grateful to Sekhar Chivukula for helpful conversations. The work of DB is supported by the Science and Engineering Research Board (SERB), Government of India grant MTR/2022/000575. DS is supported in part by the Natural Sciences and Engineering Research Council of Canada (NSERC). AD would like to acknowledge the hospitality of UCSD where the work was finalised.

\appendix

\section{Review of Inflation Formulae}
\label{app:review}

In this appendix we review the standard formulas for slow roll inflation and its connection to cosmological observations that we have used in this work 
(see for example~\cite{Baumann:2009ds}). 
Given an inflaton $\phi$ with its potential $V(\phi)$, the slow roll parameters as 
\begin{align}
    \epsilon_v &= \frac{\widetilde{M}^2_{\rm pl}}{2}\left.\left(\frac{V^\prime}{V}\right)^2\right|_{\phi=\phi_i}, \nonumber \\
    \eta_v &= \widetilde{M}^2_{\rm pl}\left.\frac{V^{\prime \prime}(\phi)}{V(\phi)}\right|_{\phi=\phi_i} \\
    \xi^2 &= \widetilde{M}^4_{\rm pl}\left.\frac{V^{\prime \prime \prime}(\phi)V^\prime(\phi)}{V^2(\phi)}\right|_{\phi=\phi_i} \nonumber 
\end{align}
where $\phi_i$ is the point in the field space where the cosmological scales leave the horizon and $\widetilde{M}_{\rm pl} = 1/\sqrt{8\pi G_N} \sim 2.4\times 10^{18}$ GeV is the reduced Planck mass.  
Now, redefining the parameters as : 
\begin{align}
    V^\prime &= \frac{1}{\phi_0}\frac{\partial V}{\partial (\phi/\phi_0)} = \frac{1}{\phi_0} V_u \quad u:=\phi/\phi_0 
\end{align}
where $\phi_0$ is the scale of the inflection point. 
Now, in terms of the redefined parameter the slow parameters are
\begin{align}
    \epsilon_v &= \left.\frac{1}{2}\left(\frac{\widetilde{M}_{\rm pl}}{\phi_0}\right)^2\left(\frac{V_u}{V}\right)^2\right|_{u=u_i} ;\nonumber \\
    \eta_v &= \left.\left(\frac{\widetilde{M}_{\rm pl}}{\phi_0}\right)^2\left(\frac{V_{uu}}{V}\right)\right|_{u=u_i}; \nonumber \\
    \xi^2 &= \left.\left(\frac{\widetilde{M}_{\rm pl}}{\phi_0}\right)^4\frac{V_{uuu}V_u}{V^2}\right|_{u=u_i} 
    \label{eq:slowRollParams}
\end{align}
Slow roll is maintained while all of these parameters are small. 

In the slow-roll regime, the above parameters can be mapped into observables that are measured in the CMB. The spectra for the scalar and the tensor perturbations are approximated in power laws given as 
\begin{align}
    \mathcal{P}_{\mathcal{R}} &= A_s(k/k_*)^{n_s-1+\frac{\alpha_s}{2}\ln k/k_*} \nonumber \\
    \mathcal{P}_{t} &= A_t(k/k_*)^{n_t}
\end{align}
where, $n_s$ is the scalar index, $\alpha_s = dn_s/d\ln k$ is the running of the scalar spectral index, $n_t$ is the tensor spectral index; $k_* = 0.05$ Mpc$^{-1}$ is the pivot scale. The tensor spectral index is related to the scalar index by the ratio $r \equiv A_t/A_s$ by $n_t \simeq -r/8$ for single field slow-roll inflation. 
Now the mapping of the above observables with the potential parameters are given as
\begin{align}
    n_s &\approx 1 - 6\epsilon_v + 2\eta_v \nonumber \\
    r &\approx 16 \epsilon_v \nonumber \\
    \alpha &\approx 16\epsilon_v \eta_v-24\epsilon_v^2 -2\xi^2 \nonumber \\
    A_s &\approx \frac{1}{12\pi^2}\frac{V^3}{V^{\prime2}}
    \label{eq:inflationObservables}
\end{align}
which are measured or bounded by CMB observations~\cite{Planck:2018jri,ACT:2020gnv}, and whose precision will be improved by future observations~\cite{SimonsObservatory:2018koc}.

The number of {\it e}-folds of inflation can be computed as 
\begin{align}
    N_e &= \frac{1}{\widetilde{M}^2_{\rm pl}}\int^{\phi_i}_{\phi_e}\frac{V(\phi)}{V^\prime(\phi)}d\phi \simeq 50-60 \; ,
\end{align}
where $\phi_e$ is the scale where the inflaton is no longer slowly rolling and inflation ends. When we scale the potential w.r.t $\phi_0$ and use $u\equiv \phi/\phi_0$, the number of {\it e}-folding is given as 
\begin{align}
    N_e &= \frac{\phi^2_0}{\widetilde{M}^2_{\rm pl}}\int^{u_i}_{u_e}\frac{V(u)}{V_u(u)}du \; ; \; V^\prime(\phi) = \frac{1}{\phi_0}V_u(u) 
    \label{eq:Ne}
    \end{align}
Using the potential given in eq.~\eqref{eq:potp}, we can reparameterize in terms of $u$ as 
\begin{align}
    \widetilde{V}(u) &= -\frac{a}{4}u^4\left[a_0 + c_0 \ln (u) + b_0 \ln^2(u)\right] \nonumber \\
    &+ aV_0 \; .
\end{align}
Defining $u_M = M/\phi_0$, the parameters of the potential can be defined piece-wise: for $u<u_M$
\begin{align}
    a_0 &= 1; \quad b_0 = b ;\nonumber \\ c_0 &= 0
\end{align}
and for $u>u_M$
\begin{align}
    a_0 &= 1 - c\ln^2(u_M); \quad b_0 = b-c ; \nonumber \\ c_0 &= 2c\ln(u_M).
\end{align}
The above number of $e$-folds in Eq.~\eqref{eq:Ne} can be re-written as 
\begin{align}
    N_e = \frac{\phi^2_0}{\widetilde{M}^2_{\rm pl}}\left[\int^{u_i}_{u_M}\frac{\widetilde{V}(u)}{\widetilde{V}_u(u)}du + \int^{u_M}_{u_e}\frac{\widetilde{V}(u)}{\widetilde{V}_u(u)}du\right].
\end{align}
Each integral can then be presented in terms of exponential integral function $\mathcal{E}(x)$ as follows:
\begin{widetext}
\begin{align}
    \mathcal{H}(a_0,b_0,c_0,u) &\equiv \int \frac{\widetilde{V}(u)}{\widetilde{V}_u(u)}du \nonumber \\
    &= \frac{1}{16\alpha}\left[2\alpha u^2 + 32V_0\left(e^{\beta_+}\mathcal{E}(-\beta_+-2\ln(u)) - e^{\beta_-}\mathcal{E}(-\beta_- - 2\ln(u))\right) \right. \nonumber \\
    &+ \left.(b_0 - \alpha)e^{-\beta_-}\mathcal{E}(\beta_- + 2\ln(u)) - (b_0 + \alpha)e^{-\beta_+}\mathcal{E}(\beta_+ + 2 \ln(u))\right] \\
    \alpha &= \sqrt{4c^2_0 + b^2_0 - 16b_0a_0}\quad , \quad \beta_\pm = \frac{1}{2} + \frac{c_0}{b_0} \pm \frac{\alpha}{2b_0} \\
    \mathcal{E}(z) &\equiv -\int^\infty_{-z}\frac{e^{-t}}{t}dt \; .
\end{align}
\end{widetext}
From this expression, the number of $e$-folds can be written as
\begin{align}
    N_e &= \frac{\phi^2_0}{\widetilde{M}^2_{\rm pl}}\left[\mathcal{H}(1,b,0,u_i)-\mathcal{H}(1,b,0,u_M) \right.\nonumber \\
    &+ \mathcal{H}(1-c\ln^2(u_M),b-c,2c\ln(u_M),u_M) \nonumber \\
    &- \left.\mathcal{H}(1-c\ln^2(u_M),b-c,2c\ln(u_M),u_e)\right] \; .
\end{align}
Along with that the inflation scale which is not at inflection point but very close to inflection point which is estimated from the spectral index $n_s|_{u=u_i} = 0.9691 \pm 0.0041$. And the end of the inflation is determined by calculating the slow-roll parameter $\epsilon_v|_{u=u_e} = 1 $.

\begin{figure}[!h]
    \centering
    \begin{tabular}{lcr}
    \includegraphics[scale=0.5]{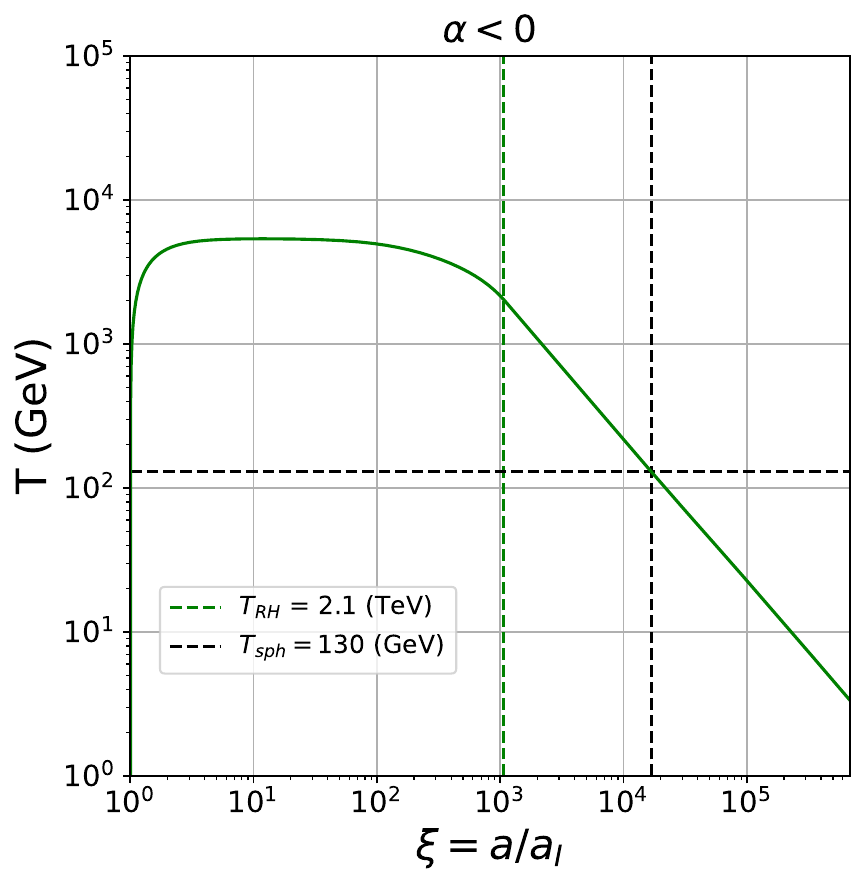}&
    \includegraphics[scale=0.5]{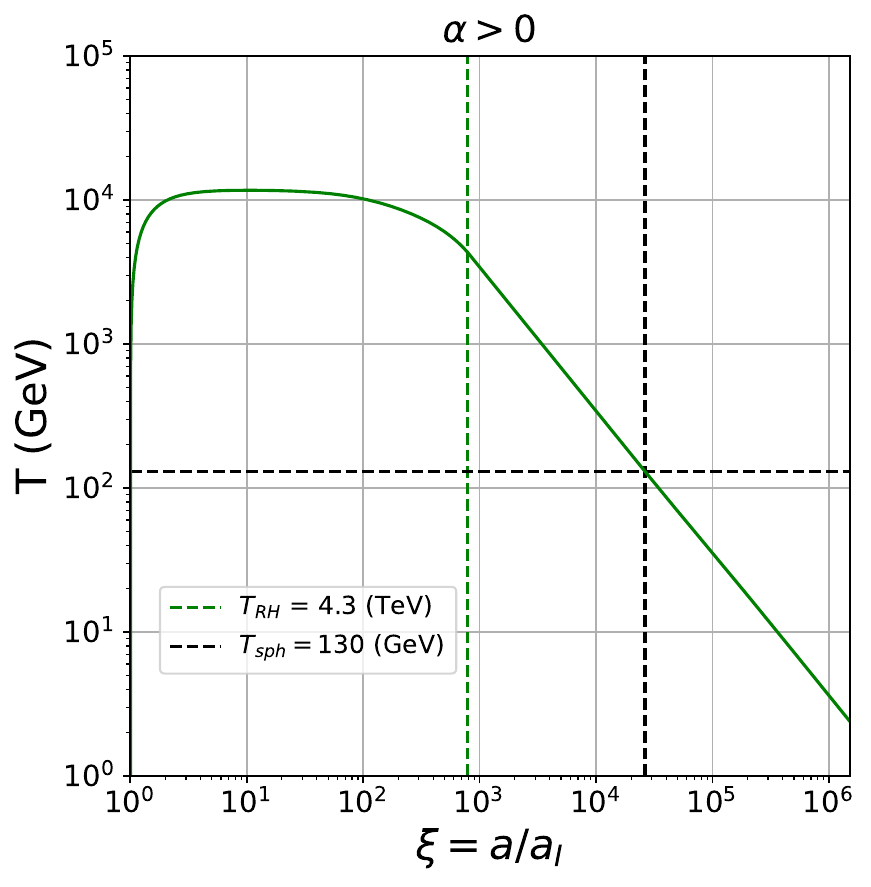}
    \end{tabular}
    \caption{Evolution of temperature with scale factor. }
    \label{fig:temp}
\end{figure}

After inflation ends, the inflaton decays its energy to a thermal bath, which in the minimal scenario is dominated by Standard Model fields. The reheating temperature can be estimated, in the instantaneous reheating approximation, as
\begin{equation}
  T_{\text{reheat}} \approx 0.2 \sqrt{\Gamma_\phi \, \widetilde{M}_{\rm pl}} \, , 
  \label{eq:reheat}
    \end{equation}
where $\Gamma_\phi$ the total decay width of the inflaton. This formula does assume that near the minimum the potential is approximately quadratic, and that is the case for the potential here. While we consider the inflaton potential to be quadratic during the oscillation, justifying the redshifting of inflaton condensate's energy as ordinary matter, we calculate the reheating temperature explicitly from the Boltzmann equations. It is the temperature at which radiation energy density overtakes that of the inflaton condensate and temperature starts redshifting as inverse of the scale factor. The evolution of bath temperature for the two benchmark points is shown in Fig. \ref{fig:temp} indicating the reheating temperature. The dashed horizontal line corresponds to the reheating temperature at which radiation energy density starts dominating and temperature redshifts as $T \propto 1/a$. The approximate expression for the reheat temperature does agree well with our numerical calculation.

While inflaton decays dominantly into right handed neutrinos thereby reheating the universe, a small fraction of it also decays into dark matter at the radiative level. The one-loop decay width of the inflaton into dark matter is given at leading order by
\begin{align}
    \Gamma_{\phi \rightarrow \psi \psi} &= \frac{\tilde{y}^2_{\phi \psi \psi}m_\phi}{8\pi}\left[1-\frac{4m^2_\psi}{m^2_\phi}\right]^{3/2}, \nonumber \\
    \tilde{y}_{\phi \psi \psi} &= g^4_D \frac{m_\psi v_\phi}{m^2_\phi-4m^2_\psi}\left[(m^2_\phi - m^2_{W_D} - 2m^2_\psi)C_0(m^2_\psi,m^2_\psi,m^2_\phi - 2m^2_\psi,m^2_\psi,m^2_{W_D},m^2_{W_D}) \right. \nonumber \\
    & \left.+ 2(B_0(m^2_\phi,m^2_{W_D},m^2_{W_D}) - B_0(m^2_\psi,m^2_\psi,m^2_{W_D}))\right],
\end{align}
where $C_0$ and $B_0$ are the Passarino-Veltman \cite{Passarino:1978jh} scalar integrals given by
\begin{align}
    C_0(p^2_1,p^2_2,2p_1.p_2,m^2_1,m^2_2,m^2_3) &= \frac{(2\pi\mu)^{4-D}}{i\pi^2}\int d^Dl\frac{1}{(l^2-m^2_1)((l-p_1)^2-m^2_2)((l+p_2)^2-m^2_3)}, \nonumber \\
    B_0(p^2,m^2_1,m^2_2) &= \frac{(2\pi\mu)^{4-D}}{i\pi^2}\int d^Dl\frac{1}{(l^2-m^2_1)((l+p_1)^2-m^2_2)}.
\end{align}
The expression for $\tilde{y}_{\phi \psi \psi}$ reduces to the one in Eq.~\eqref{eq:inflatonDMdecay} in the large $m_{W_D}$ limit.

\bibliographystyle{apsrev4-1}
\bibliography{ref, ref_db}
\end{document}